\documentclass[11pt]{article}
\usepackage[sc]{mathpazo}
\usepackage{fullpage}
\usepackage[authoryear,sectionbib,sort]{natbib}
\usepackage{graphicx}
\usepackage{adjustbox}
\usepackage[boxruled,linesnumbered]{algorithm2e}
\usepackage{multirow}
\usepackage{amsmath}
\usepackage[utf8]{inputenc}
\usepackage{lineno}
\usepackage{titlesec}
\usepackage{graphicx}
\usepackage{subcaption}
\usepackage{booktabs,multirow}
\usepackage{here}
\usepackage{url}
\usepackage{siunitx}

\newcommand{\argmax}[1]{\underset{#1}{\operatorname{arg}\,\operatorname{max}}\;}

\titleformat{\section}[block]{\Large\bfseries\filcenter}{\thesection}{1em}{}
\titleformat{\subsection}[block]{\Large\itshape\filcenter}{\thesubsection}{1em}{}
\titleformat{\subsubsection}[block]{\large\itshape}{\thesubsubsection}{1em}{}
\titleformat{\paragraph}[runin]{\itshape}{\theparagraph}{1em}{}[. ]

\title{Forecasting insect abundance using time series embedding and machine learning}

\author{Gabriel R. Palma$^{1, 6,\ast}$ \and 
Rodrigo F. Mello$^{2}$ \and
Wesley A.C. Godoy$^{3}$ \and
Eduardo Engel$^{3}$ \and
Douglas Lau$^{4}$ \and
Charles Markham$^{1, 5}$ \and
Rafael A. Moral$^{1, 6}$}

\date{}

\begin{document}

\maketitle

\noindent{} 1. Hamilton Institute, Maynooth University, Maynooth, Ireland;

\noindent{} 2. DOTI, Itaú Unibanco SA, S\~{a}o Paulo, Brazil

\noindent{} 3. Department of Entomology and Acarology, University of S\~{a}o Paulo, Piracicaba, Brazil;

\noindent{} 4. Brazilian Agricultural Research Corporation (Embrapa Trigo), Passo Fundo, Rio Grande do Sul, Brazil;

\noindent{} 5. Department of Computer Science, Maynooth University, Maynooth, Ireland;

\noindent{} 6. Department of Mathematics and Statistics, Maynooth University, Maynooth, Ireland;

\noindent{} $\ast$ Corresponding author; e-mail: gabriel.palma.2022@mumail.ie

\bigskip


\bigskip


\bigskip

\textit{Manuscript type}: Research paper. 

\bigskip

\noindent{\footnotesize Prepared using the suggested \LaTeX{} template for \textit{Am.\ Nat.}}

\newpage{}

\section*{Abstract}

Implementing insect monitoring systems provides an excellent opportunity to create accurate interventions for insect control. However, selecting the appropriate time for an intervention is still an open question due to the inherent difficulty of implementing on-site monitoring in real-time. This decision is even more critical with insect species that can abruptly increase population size. A possible solution to enhance decision-making is to apply forecasting methods to predict insect abundance. However, another layer of complexity is added when other covariates are considered in the forecasting, such as climate time series collected along the monitoring system. Multiple possible combinations of climate time series and their lags can be used to build a forecasting method. Therefore, this research paper proposes a new approach to address this problem by combining statistics, machine learning, and time series embedding. We used two datasets containing a time series of aphids and climate data collected weekly in Coxilha and Passo Fundo municipalities in Southern Brazil for eight years. We conduct a simulation study based on a probabilistic autoregressive model with exogenous time series based on Poisson and negative binomial distributions to check the influence of incorporating climate time series on the performance of our approach. We pre-processed the data using our newly proposed approach and more straightforward approaches commonly used to train machine learning algorithms in time series problems. We evaluate the performance of the selected machine algorithms by looking at the Root Mean Squared Error obtained using one-step-ahead forecasting. Based on Random Forests, Lasso-regularised linear regression, and LightGBM regression algorithms, our novel approach yields competitive forecasts while automatically selecting insect abundances, climate time series and their lags to aid forecasting.

\newpage{}

\section{Introduction} 

Insect outbreaks have frequently been documented in pest populations \citep{Lynch2009, Santos2017, Lynch2018}. This ecological disturbance can affect forests and agroecosystems \citep{Wallner1987, Nair2001, Lantschner2019}, resulting in economic and environmental damage. Yield loss is an example of the impacts of the insect outbreak due to the consumption of plants by the pest species. Arthropod pests are responsible for $20\%$ of global annual crop losses \citep{mateos2022}. Also, the transmission of diseases from insect-plant interaction \citep{perring1999, smyrnioudis2001, brown2002, heck2018, hoffmann2023} provides a clear example of how vital the correct management of insect outbreak is for avoiding economic damage. 

When insect outbreaks start evolving in natural regions, they can disturb the food chain in the ecosystem, impacting biodiversity by reducing the population of other essential species \citep{muller2008}. There are several examples of insect pest species related to outbreaks in forests \citep{singh2009} and crops, such as \textit{Thyrinteina arnobia} and \textit{Stenalcidia sp} (Geometridae) \citep{Zanuncio2006}, \textit{Oncideres impluviata} (Cerambycidae) \citep{Ono2014}, \textit{Chrysodeixis includes} and \textit{Anticarsia gemmatalis} (Lepidoptera: Noctuidae) \citep{ muller2008, Bueno2010, Santos2017, palma2023}. 

These examples motivate developing and implementing forecasting methods to prevent insect outbreaks. Especially methods that predict the best moment to proceed with interventions in the pest population. The possible types of interventions are well-studied in the Integrated Pest Management (IPM) field. So, the vast number of alternative solutions, such as biological, physical, chemical and biotechnology control approaches \citep{mateos2022} provides an excellent toolkit for growers; however, methods supporting accurate decisions for preventing insect outbreaks still lead to open research questions, helping the discovery of the appropriate solution to be applied in the field to reduce crop losses. 

There are several approaches to implementing outbreak forecasting. One alternative is to predict the event using an algorithm to classify a binary problem (outbreak and non-outbreak) or provide forecasts for the abundance of insects. Based on event prediction, machine learning algorithms have demonstrated high performance for classifying insect outbreaks~\citep{ramazi2021, palma2023}. Also, machine learning has demonstrated promising results for forecasting insect abundance~\citep{scavuzzo2018, chen2019, zhao2020, rouabah2022, ceia2023, kishi2023}. The findings have shown that in some cases, machine learning methods achieve higher performance than traditional methods when analysing different types of time series \citep{khedmati2020, spiliotis2020, buttner2021, hamdoun2021, maaliw2021, masini2023}. These algorithms can include multiple exogenous time series to obtain forecasts of a target time series, and recent reviews highlighted the efficiency of high-dimension algorithms, such as Lasso-type, Random Forests and Ensemble-based algorithms~\citep{masini2023}. These examples inspire the application of machine-learning algorithms to insect abundance forecasting in more depth. 

Climate covariates are also collected over time in many insect monitoring systems. Since insects are poikilothermic organisms subject to meteorological variations on different temporal scales, climate coraviates may play an important role in their development. Meteorological variables such as temperature and rainfall are among the main abiotic factors that influence insect population dynamics. Among the complex pests affected by short- and long-term climate changes, aphids are the most sensitive and are commonly used as a study model \citep{engel2022}. This brings additional opportunities to enhance insect abundance forecasting. However, it introduces more complexity to the analysis, considering the multiple combinations of lags from each climate time series \citep{brabec2014} that can be used as features of machine learning algorithms. To address this problem, this paper introduces a novel approach for predicting insect abundance by combining statistics, machine learning, and time series analysis techniques. We introduce a framework for understanding the causal effects of climate on insect abundance, considering multiple lags. Then, we use LightGBM, Lasso-regularised linear regression, and Random Forest algorithms to predict crop pest dynamics based on the forecast-focussed causal analysis. Finally, we combine these techniques for predicting insect abundance, illustrating our findings with two real-time series of aphid populations in the State of Rio Grande do Sul (RS) in Southern Brazil, and a simulation study based on Poisson and negative binomial autoregressive models with exogenous time series.

\section{Methods}

\subsection{Reconstructing time series dependencies}
\label{TimeSeriesReconstruction}
Before proceeding with any learning strategy applied to temporal data, we must reconstruct it by unfolding time dependencies among observations~\citep{Mello2018}. For instance, classical time series modelling approaches such as AR, ARMA, and ARIMA perform such reconstruction implicitly while modelling the influences that past observations have on current ones. In this context, we apply Takens' embedding theorem to explicitly reconstruct each observation $x(t)$ from a time series $X$ with $T$ observations, for all $t = 1, \ldots, T$, in a phase space coordinates $\mathbf{\Phi}$ in the form:
\begin{align}\label{embeddingdim}
\phi_t = (x(t), x(t + \tau), \ldots, x(t + (\tau m - \tau))),
\end{align}
having $m$ as the embedding dimension or the number of spatial axes, and $\tau$ as the time delay in between consecutive observations, finally $\phi_t$ corresponds to a position vector or state in a phase space $\mathbf{\Phi}$, i.e., $(\phi_1, \phi_2, \ldots, \phi_{T-(\tau m - \tau)}) \in \mathbf{\Phi}$.

In our particular scenario, given our interest in analysing cause-effect relationships among time series observations, we employed Granger's causality~\citep{pearl2000, shojaie2022granger} to map how a given exogenous or explanatory variable (another time series such as temperatures, rainfall, etc.) influences or anticipates events on the target time series (e.g. population size of insects) which is based on a set of past observations with the time delay $\tau = 1$. The embedding dimension $m$ can be estimated using Autoregressive models (AR)~\citep{box2015}. In addition to such reconstruction, Granger's method requires the following additional steps:

\begin{enumerate}
    \item Take the target time series $Y$ (insect abundances over time) and employ first-order differences while it contains a relevant non-stationary component, which can be detected based on the Augmented Dickey-Fuller Test;
    \item Take every exogenous time series $X_i$, where $i = 1, \ldots, I$ and $I$ is the number of exogenous times series. For each $X_i$, ensure it is stationary by performing the same steps considered in the previous item;
    \item Employ the cross-correlation function ($CCF$) on every pair on the stationary versions of an exogenous $X_i$ and the target time series $Y$ to measure the time delay for which the exogenous series has the greatest correlation (maximal correlation $MC(X_i, Y) = \argmax{CCF(X_i, Y)}$) with a future observation of the target time series;
    \item Take time delay $\tau = 1$ and estimate the embedding dimension $m$ by using the AR model on every time series (exogenous and target); 
    \item Employ Takens' embedding theorem to reconstruct all time series (exogenous and target), resulting in one data frame or panel per series;
    \item The maximal correlation $MC(X_i, Y)$ is then used as a criterion to join data frames into a single dataset $\mathcal{D}$ and perform learning.
\end{enumerate}

It is worth detailing how those data frames are merged based on the maximal correlation of every exogenous time series $X_i$ with the target series $Y$ described as $MC(X_i, Y)$. Suppose $MC(X_\text{rainfall},Y)=-5$, this means every current value of $Y$, this is $Y(t)$, is most likely to depend on $X_\text{rainfall}(t-5)$. Consequently, they should be aligned before proceeding with modelling.

Suppose $m_\text{rainfall}=3$ and $m_\text{target} = 2$, then we will have data frames built as follows: i) every row of the rainfall data frame will contain $(X_\text{rainfall}(t-2), X_\text{rainfall}(t-1), X_\text{rainfall}(t))$; and ii) every row of the target data frame will contain $(Y(t-1), Y(t))$. Now, taking into consideration $MC(X_\text{rainfall}, Y)=-5$, we must merge those data frames to produce $(X_\text{rainfall}(t-7), X_\text{rainfall}(t-6), X_\text{rainfall}(t-5), Y(t-1), Y(t))$ in which the three columns associated with the exogenous series were time displaced so that $X_\text{rainfall}(t-5)$ is used to bring as much information as possible to predict $Y(t)$. The same steps must be performed on all exogenous variables to obtain a single data frame $\mathcal{D}$, plugging all explanatory time series into our target series. Any machine learning method can be used to forecast $Y(t)$ after combining all explanatory time series into the target series based on the described method. Here, we present forecasting results based on three algorithms commonly used for this context: LightGBM, Random Forests and Lasso-regularised linear regression.

\subsection{Insect time series datasets}

To compare the performance of the selected algorithms against each other, we used two real datasets of $211$ observations each, including the total number of sampled aphids obtained from a monitoring system in Coxilha (S\ang{28;11;16.9} W\ang{52;19;31.7}) and Passo Fundo (S\ang{28;13;36.6} W\ang{52;24;13.4}) in the State of Rio Grande do Sul in Southern Brazil~\citep{engel2022}. Each observation of the insect abundance time series contains climate covariates. In Table~\ref{features}, we present a sample from the time series collected over the weeks related to the datasets. Both regions use the same climate time series, given that the sampling areas are separated by approximately 12 kilometres. Thus, the main difference is the aphids' population dynamics.

\begin{table}[ht]
\caption{A sample of $4$ weeks showing the features collected for Coxilha and Passo Fundo regions. The dataset contains the region (\textit{region}), year (\textit{year}), week (\textit{w}), the temperature (\textit{tmin} and \textit{tmax}), rainfall (\textit{pmm}), relative humidity (\textit{ur}), wind speed (\textit{wmax} and \textit{wmean}), the temperature at $5$ and $10$ cm of the soil (\textit{st5cm} and \textit{st5cm}), and the aphid community total abundance (\textit{aphids}).}
	
\centering
\begin{adjustbox}{max width=\textwidth}
\begin{tabular}{crrrrrrrrrrrr}
  \hline
\textit{region} & \textit{year} & \textit{W} & \textit{tmax} & \textit{tmin} & \textit{tmean} & \textit{pmm} & \textit{Ur} & \textit{wmax} & \textit{wmean} & \textit{st5cm} & \textit{st10cm} & \textit{aphids} \\ 
  \hline
  Coxilha &2015 & 1 & 28.03 & 18.35 & 22.27 & 162.00 & 79.10 & 12.31 & 3.12 & 7.09 & 25.96& 102 \\ 
  Coxilha &2015 & 2 & 30.80 & 19.30 & 23.48 & 33.60 & 78.00 & 14.12 & 4.20 & 6.53 & 26.80& 105 \\ 
  Coxilha &2015 & 3 & 26.94 & 18.03 & 21.67 & 114.10 & 84.00 & 11.97 & 2.81 & 4.87 & 24.69& 23 \\ 
  Coxilha &2015 & 4 & 28.24 & 17.31 & 22.21 & 17.40 & 78.14 & 11.39 & 3.14 & 9.20 & 26.64& 100 \\
  
  Passo Fundo &2015 & 1 & 28.03 & 18.35 & 22.27 & 162.00 & 79.10 & 12.31 & 3.12 & 7.09 & 25.96& 0 \\ 
  Passo Fundo &2015 & 2 & 30.80 & 19.30 & 23.48 & 33.60 & 78.00 & 14.12 & 4.20 & 6.53 & 26.80& 0\\ 
  Passo Fundo &2015 & 3 & 26.94 & 18.03 & 21.67 & 114.10 & 84.00 & 11.97 & 2.81 & 4.87 & 24.69& 0\\ 
  Passo Fundo &2015 & 4 & 28.24 & 17.31 & 22.21 & 17.40 & 78.14 & 11.39 & 3.14 & 9.20 & 26.64& 12\\
  
   \hline
   \end{tabular}
   \end{adjustbox}
   \label{features}
\end{table}

\subsection{Simulation study}
The simulation study aims to gather insights into the requirements of the proposed method. Specifically, we aim to determine if the presence of climate time series impacts the performance of the proposed method when analysing the target time series. To conduct the simulation study, we begin by estimating the parameters of an Autoregressive model with exogenous time series (ARX) based on the Passo Fundo region dataset, including the insect abundance and climate time series. Let $y(t)$ be an observation from a discrete time series of insect densities $Y$, $x_i(t)$ be an observation from a time series of climate features $X_i$, where $i = \{ 1, \ldots, 9\}$ (i.e. the climate variables presented in Table ~\ref{features}). The autoregressive model of order $p$ with exogenous time series can be written as: 
\begin{align}
    \label{ARX}
    \nonumber
    y(t) &= C + \omega_1 y(t-1) + \omega_2 y(t-2) +  \ldots + \omega_p y(t-p)~+\\ 
     &~~~~ \theta_{1}B^{p}x_1(t) + \theta_{2}B^{p}x_2(t) + \ldots + \theta_{9}B^{p}x_9(t)~+ \\ \nonumber
     &~~~~ \theta_{10}x_1(t) + \theta_{11}x_2(t) + \ldots + \theta_{18}x_9(t) + \epsilon(t), \nonumber
\end{align}
where $\epsilon(t) \sim N(0, \sigma^2)$, $B^{p}x_i(t) = x_i(t - p)$ is the backshift operator, $\Omega = (\omega_1, \ldots, \omega_p)^\top$ are the autoregressive coefficients for the target time series $Y$, $\Theta = (\theta_1, \ldots, \theta_{18})^\top$ are the coefficients related to each climate time series $X_i$ and their lags. To estimate the coefficients of the presented model, we first transformed the time series of insect densities by taking $Y^{\prime} = \log(Y+0.1)$. Then, all the coefficients were estimated using the package \texttt{fable}~\citep{fable2023} for R software~\citep{Rsoftware}. 

To explore different scenarios of insect dynamics, we used two distributions for generating $Y$ and $Y^{\prime}$. For the first scenario, we assumed that $Y(t+1)~\sim \mbox{Poisson}(\lambda = \exp(y^{\prime}(t)-0.1))$. For the second distribution we assumed that $Y(t+1)~\sim \mbox{Negative binomial}(\lambda = \exp(Y^{\prime}(t)-0.1), \theta = 1.8)$ to introduce overdispersion to the simulation. To investigate the impact of climate time series on the method performance, we first tested a scenario where we do not estimate any of the $\Theta$ parameters, representing no influence of climate time series on the simulated insect abundance. For the second scenario, we estimated $\theta_1, \theta_2, \theta_3, \theta_4, \theta_5$, $\theta_{10}, \theta_{11}, \theta_{12}, \theta_{13}, \theta_{14}$ (related to \textit{tmax}, \textit{tmin}, \textit{tmean}, \textit{pmm}, \textit{ur} climate time series and their lags $p$), representing the influence of half of the climate time series on the simulated insect abundance. Finally, for the third scenario, we estimated all $\Theta$ parameters related to the climate time series, representing the influence of all climate time series and their lags $p$ on the simulated insect abundance. We used for each scenario $p = \{1, 3, 5\}$. For each combination of scenario and value of $p$, we generated $50$ time series for both distributions, Poisson and negative binomial. Also, each simulated time series had $211$ observations.

\subsection{Forecasting performance of machine learning algorithms}

To compare the performance of our novel method, we (i) used only the target time series (population of insects), with up to $3$ or $6$ lags behind to forecast future observations, (ii) used all climate time series as predictors with no lags, and (iii) all climate time series with target time series (population of insects) up to $3$ or $6$ lags behind to forecast future observations. We performed validation by obtaining one-step ahead forecasts for the entire time series (apart from the first $30$ and $60$ observations used for training the learning algorithms). For all time series of the case and simulation study, we trained Random Forests, Lasso-regularised linear regression and LightGBM algorithms for each approach (including all climate time series with no lags, time series reconstruction, taking the naive approach with up to $3$ or $6$-step lagged target series, and all climate times with up to $3$ or $6$-step lagged target series) and obtained their performance based on the Root Mean Squared Error (RMSE). 

In addition, the proposed method provides a data frame $\mathcal{D}$ that selects climate, target time series and their lags during the reconstruction process, as described in Section~\ref{TimeSeriesReconstruction}. So, a data frame $\mathcal{D}$ with these features is created for every forecast during the one-step-ahead forecasting. To analyse the selection procedure that creates $\mathcal{D}$ on the forecasting performance of the proposed method, we collected the number of selected features of $\mathcal{D}$ (the sum of the number of climate, target time series and their lags) for every forecast. Also, we collected the forecasting absolute error computed by the difference between the Random Forests' prediction and observed insect abundance. Finally, we analysed the correlation between the number of selected features of $\mathcal{D}$ (the sum of the number of climate, target time series and their lags) and the absolute error of the Random Forests algorithm.

We presented the forecasts of the Random Forests algorithm due to its inherent capability to incorporate non-linear relationships, allowing us to explore non-linear associations of most features provided by $\mathcal{D}$. Lasso-regularised linear regression would solely include linear associations, and the penalty provided by the L1 regularisation would decrease the number of features used in the forecasting, impacting the visualisation of the time series selection on the forecasting performance. The Light GBM could also be used based on these points. However, considering that the Random Forests algorithm has been commonly reported in papers with entomological applications~\citep{chen2019, valavi2021modelling, masini2023, palma2023}, we solely report the results with this learning algorithm.

\section{Results}

\subsection{Case study}
\begin{figure}[!ht]\centering
\includegraphics[width=1\textwidth]{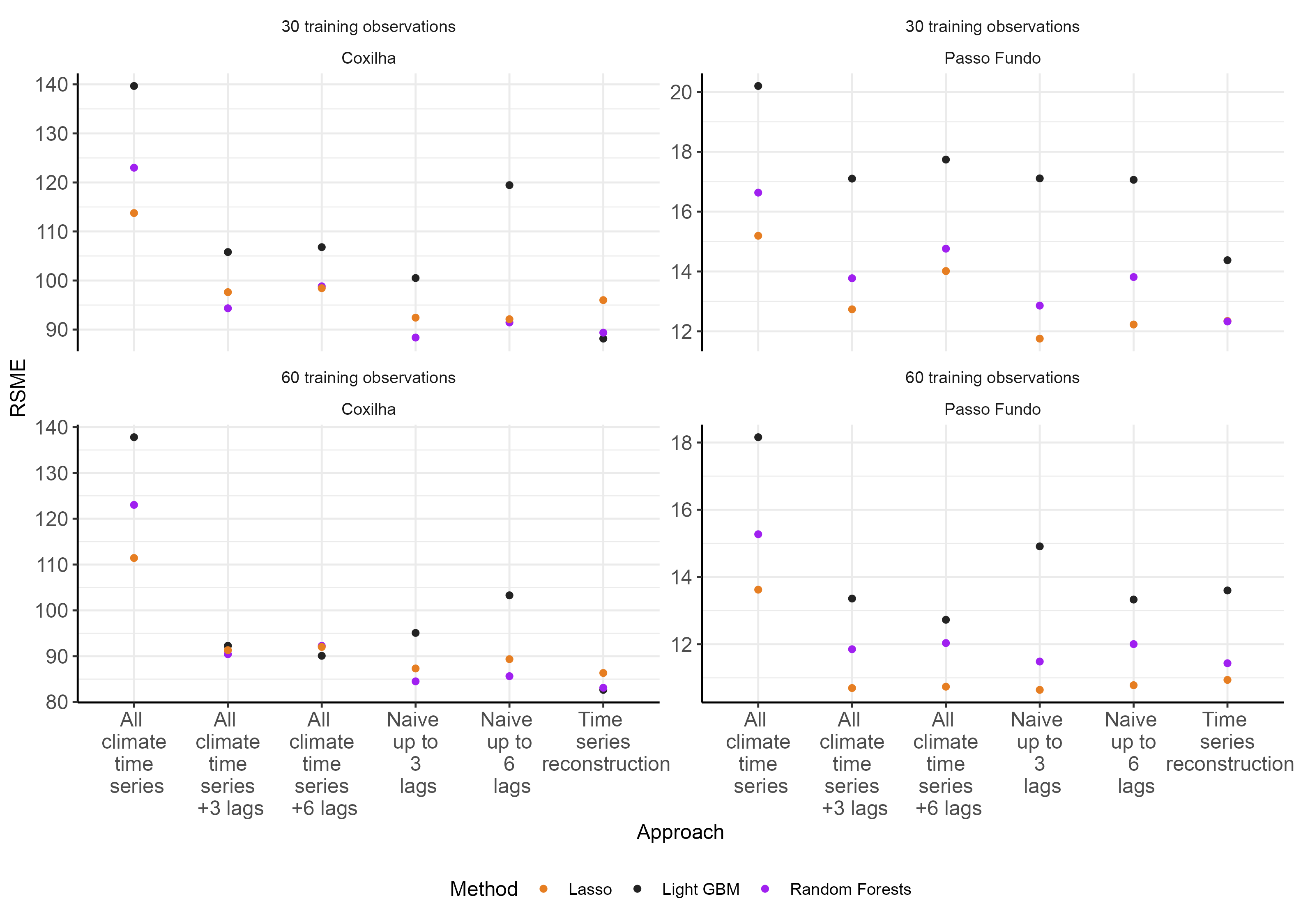}
\caption{Performance of the Random Forests, Lasso-regularised linear regression, and LightGBM algorithms for each dataset (Coxilha and Passo Fundo with aphid's abundances), approach (including all climate time series with no lags, time series reconstruction, taking the naive approach with up to $3$ or $6$-step lagged target series, and all climate times with up to $3$ or $6$-step lagged target series) and the initial number of training samples used for each learning algorithm.}
\label{case_study_performance}
\end{figure}

Figure~\ref{case_study_performance} shows, as an overall result, that as the initial number of training samples increases, the RSME reduces in most scenarios. Considering all scenarios, approaches, and learning algorithms, the average RMSE reduced from $58.4$ to $54.0$ when the initial training sample increased. Overall, the average RMSE for Random Forests, Lasso-regularised linear regression and LightGBM were $54.3$, $53.9$ and $60.5$. The average RMSE considering all datasets and algorithms for the proposed method, all climate time series with no lags, all climate time series with up to $3$ or $6$-lagged target series, and taking the naive approach with up to $3$ or $6$-lagged target series were, respectively, $50.1$, $70.6$, $54.3$, $55.0$, $52.2$, and $55.0$. 

For Lasso-regularised linear regression, the average RMSE considering all datasets for the proposed method, all climate time series with no lags, all climate time series with up to $3$ or $6$-lagged target series, and taking the naive approach with up to $3$ or $6$-lagged target series were, respectively, $51.4$, $63.5$, $53.1$, $53.8$, $50.5$ and $51.1$. For Random Forests, the average RMSE considering all datasets for the proposed method, all climate time series with no lags, all climate time series with up to $3$ or $6$-lagged target series, and taking the naive approach with up to $3$ or $6$-lagged target series were, respectively, $49.1$, $69.5$, $52.6$, $54.5$, $49.3$ and $50.7$. The same scenarios for LightGBM resulted in $49.7$, $79.0$, $57.1$, $56.8$, $56.9$, and $63.3$.

\begin{figure}[!ht]\centering
\includegraphics[width=1\textwidth]{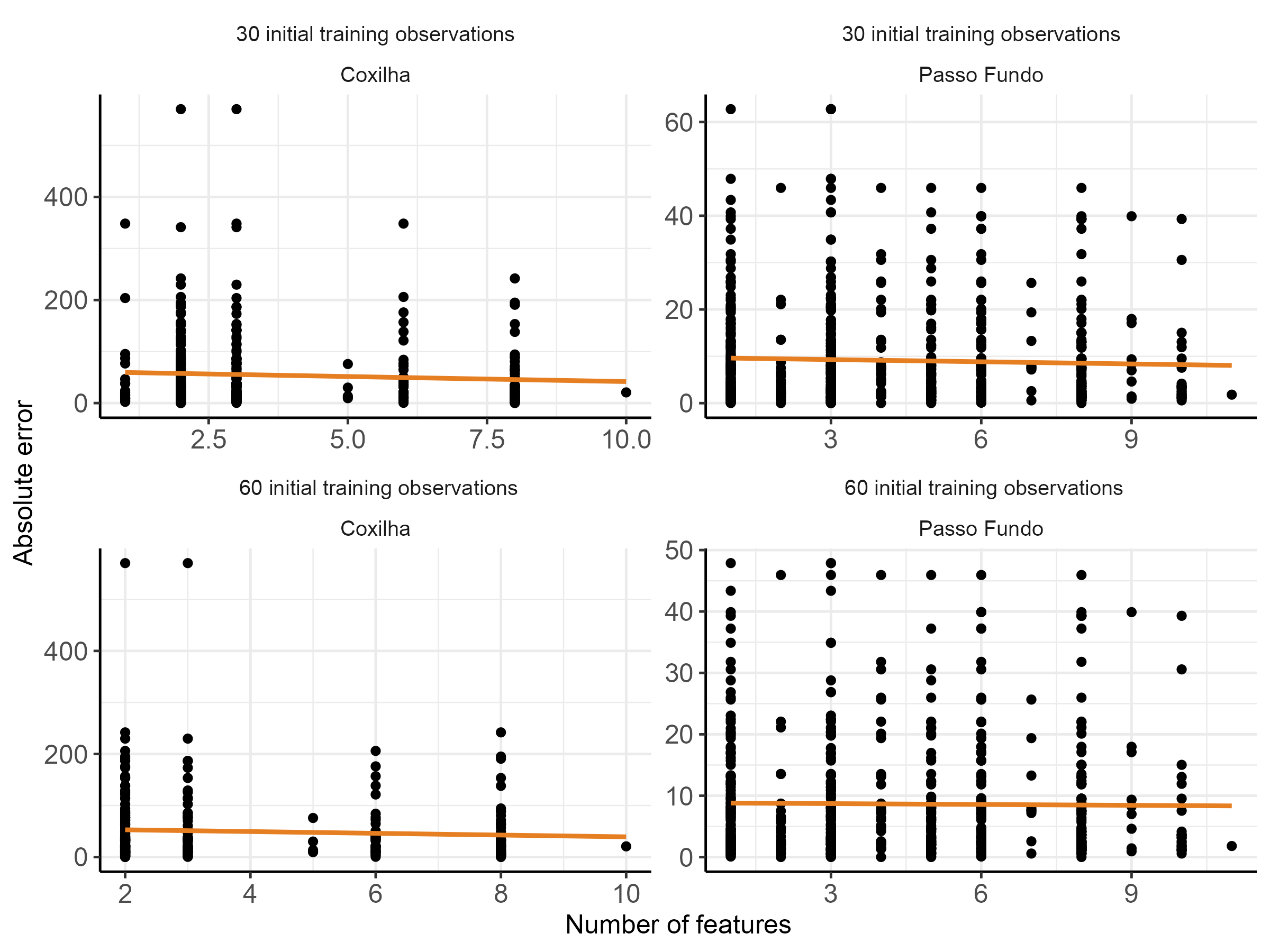}
\caption{Scatter plots of the Random Forests' forecasting absolute error and the number of selected features of $\mathcal{D}$ (the sum of the number of climate, target time series and their lags) per forecast by our approach for the datasets of Coxilha and Passo Fundo regions.}
\label{case_study_features}
\end{figure}

Figure~\ref{case_study_features} shows that for Coxilha, the obtained correlation between the number of selected features of $\mathcal{D}$ (the sum of the number of climate, target time series and their lags) based on the reconstruction approach and the absolute error of each prediction based on Random Forests for each dataset for $30$ and $60$ initial training samples are, respectively, $-0.06$ and $-0.06$. For Passo Fundo, the obtained correlation between both variables for $30$ and $60$ initial training samples are $-0.04$ and $-0.01$. 

\subsection{Simulation study}
\begin{figure}[!ht]\centering
\includegraphics[width=1\textwidth]{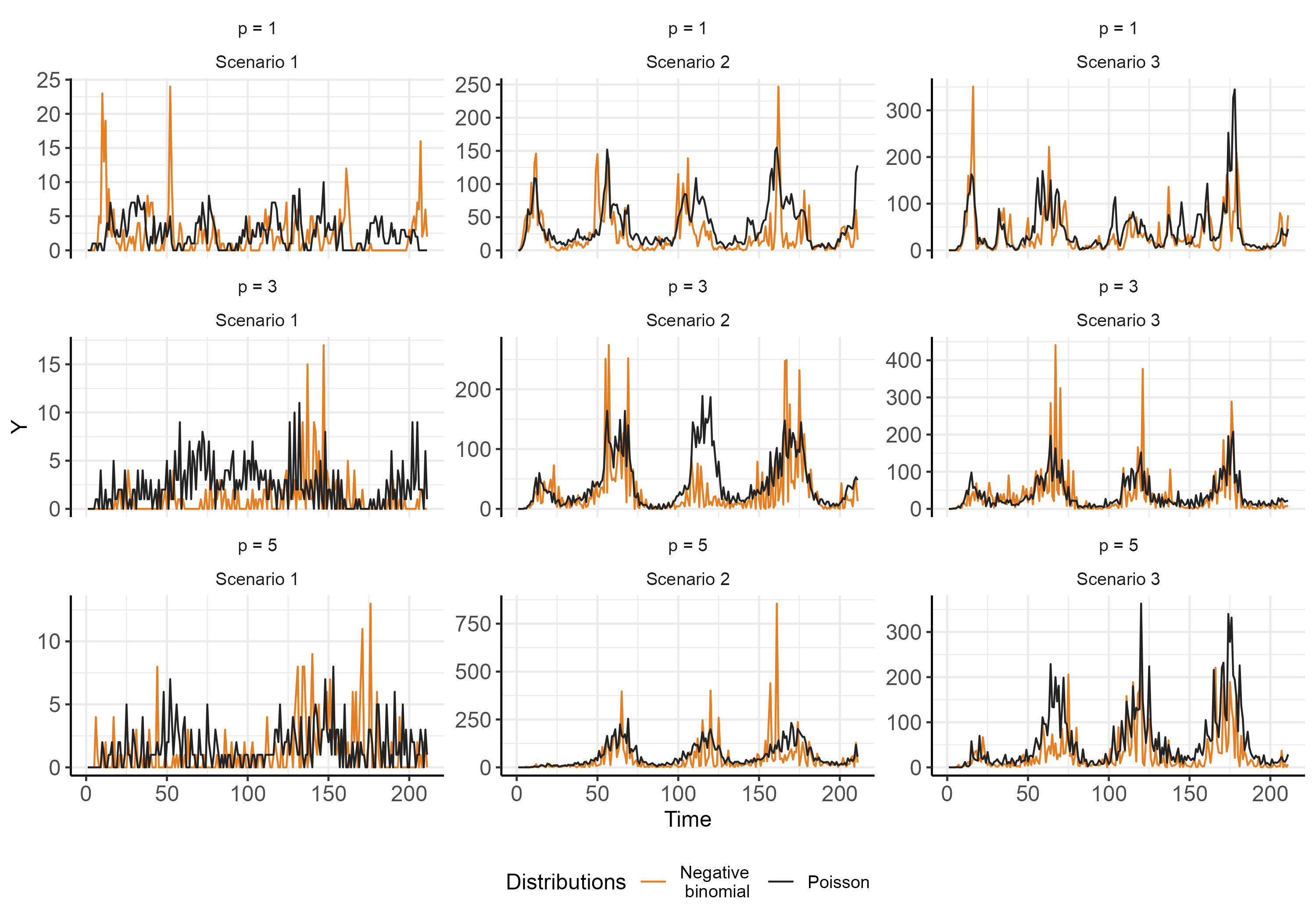}
\caption{A sample of one simulated time series built upon Poisson and negative binomial ARX considering the lags $p = \{1, 3, 5\}$ and the scenarios: $1$ - No influence of climate time series presented in Table~\ref{features} on simulated insect abundances; $2$ - influence of five climate time series presented in Table~\ref{features} on simulated insect abundances; and $3$ - Influence of all climate time series presented in Table~\ref{features} on simulated insect abundances.}
\label{simulation_time_series}
\end{figure}

Table~\ref{ARXParameters} presents the estimated parameters of the autoregressive model with exogenous time series described in Equation~\ref{ARX} and their standard errors for the simulation study. Figure~\ref{simulation_time_series} illustrates the time series generated based on the Poisson and negative binomial ARX by presenting a sample of one time series per case of the simulation study. 

\begin{table}[!ht]
\centering
\begin{adjustbox}{max width=\textwidth}
\begin{tabular}{cccccccccccccc}
\hline
Scenarios & $p$ & Intercept & $\omega_1$ & $\omega_2$ & $\omega_3$ & $\omega_4$ & $\omega_5$ & $\theta_1$ & $\theta_2$ & $\theta_3$ & $\theta_4$ & $\theta_5$ & $\theta_6$ \\
\hline

 & $1$ & $0.58$ ($0.10$) & $0.60$ ($0.06$) & - & - & - & - & - & - & - & - & - & - \\
$1$ & $3$ & $0.49$ ($0.10$) & $0.50$ ($0.06$) & $0.24$ ($0.07$) & - $0.08$ ($0.07$) & - & - & - & - & - & - & - & - \\
 & $5$ & $0.51$ ($0.10$) & $0.51$ ($0.07$) & $0.23$ ($0.08$) & - $0.08$ ($0.08$) & $0.04$ ($0.08$) & $-0.04$ ($0.07$) & - & - & - & - & - & - \\[0.5cm]

 & $1$ & $0.14$ ($5.49$) & $0.59$ ($0.06$) & - & - & - & - & $-0.16$ ($0.19$) & $-0.01$ ($0.24$) & $0.18$ ($0.38$) & $-0.00$ ($0.00$) & $-0.00$ ($0.03$) & - \\
$2$ & $3$ & $-3.27$ ($4.46$) & $0.49$ ($0.07$) & $0.22$ ($0.08$) & $-0.11$ ($0.08$) & - & - & $0.15$ ($0.18$) & $0.01$ ($0.23$) & $-0.12$ ($0.36$) & $-0.00$ ($0.00$) & $0.04$ ($0.03$) & - \\
 & $5$ & $-1.55$ ($4.55$) & $0.43$ ($0.07$) & $0.22$ ($0.08$) & $-0.02$ ($0.08$) & $0.04$ ($0.08$) & $-0.04$ ($0.07$) & $0.01$ ($0.18$) & $0.03$ ($0.22$) & $0.02$ ($0.35$) & $-0.00$ ($0.00$) & $0.00$ ($0.03$) &  - \\[0.5cm]

 & $1$ & $5.61$ ($5.80$) & $0.59$ ($0.06$) & - & - & - & - & $-0.11$ ($0.20$) & $0.09$ ($0.26$) & $0.05$ ($0.41$) & $-0.00$ ($0.00$) & $-0.04$ ($0.03$) & $-0.00$ ($0.09$) \\
$3$ & $3$ & $-1.01$ ($4.86$) & $0.49$ ($0.07$) & $0.22$ ($0.08$) & $-0.13$ ($0.08$) & - & - & $-0.124$ ($0.18$) & $-0.04$ ($0.23$) & $-0.09$ ($0.36$) & $-0.00$ ($0.00$) & $0.05$ ($0.03$) & $0.04$ ($0.09$) \\
 & $5$ & $2.46$ ($5.19$) & $0.43$ ($0.08$) & $0.24$ ($0.08$) & $-0.02$ ($0.09$) & $-0.01$ ($0.08$) & $-0.044$ ($0.08$) & $-0.03$ ($0.18$) & $-0.02$ ($0.23$) & $0.01$ ($0.36$) & $-0.00$ ($0.00$) & $-0.01$ ($0.03$) & $0.06$ ($0.08$) \\
\hline
Scenarios & $p$ & $\theta_7$ & $\theta_8$ & $\theta_9$ & $\theta_{10}$ & $\theta_{11}$ & $\theta_{12}$ & $\theta_{13}$ & $\theta_{14}$ & $\theta_{15}$ & $\theta_{16}$ & $\theta_{17}$ & $\theta_{18}$ \\
\hline
 & $1$ & - & - & - & - & - & - & - & - & - & - & - & - \\
$1$ & $3$ & - & - & - & - & - & - & - & - & - & - & - & - \\
 & $5$ & - & - & - & - & - & - & - & - & - & - & - & - \\[0.5cm]

  & $1$ & - & - & - & $-0.01$ ($0.19$) & $-0.16$ ($0.25$) & $0.26$ ($0.37$) & $-0.00$ ($0.00$) & $0.01$ ($0.03$) & - & - & - & - \\
$2$ & $3$ & - & - & - & $0.01$ ($0.18$) & $0.00$ ($0.22$) & $0.06$ ($0.35$) & $-0.00$ ($0.00$) & $-0.01$ ($0.03$) & - & - & - & - \\
  & $5$ & - & - & - & $0.02$ ($0.17$) & $-0.08$ ($0.22$) & $0.13$ ($0.35$) & $-0.00$ ($0.00$) & $0.00$ ($0.03$) & - & - & - & - \\[0.5cm]
  
 & $1$ & $-0.26$ ($0.22$) & $-0.09$ ($0.09$) & $0.00$ ($0.10$) & $0.07$ ($0.19$) & $-0.06$ ($0.25$) & $0.23$ ($0.40$) & $-0.00$ ($0.00$) & $0.01$ ($0.03$) & $-0.19$ ($0.09$) & $0.00$ ($0.22$) & $0.05$ ($0.09$) & $-0.14$ ($0.12$) \\
$3$ & $3$ & $-0.21$ ($0.21$) & $0.00$ ($0.08$) & $0.07$ ($0.10$) & $0.07$ ($0.19$) & $0.12$ ($0.23$) & $0.02$ ($0.36$) & $0.00$ ($0.00$) & $-0.03$ ($0.03$) & $-0.14$ ($0.08$) & $-0.02$ ($0.20$) & $0.02$ ($0.08$) & $-0.14$ ($0.10$) \\
 & $5$ & $-0.26$ ($0.20$) & $-0.03$ ($0.08$) & $0.15$ ($0.10$) & $0.09$ ($0.18$) & $0.07$ ($0.23$) & $0.02$ ($0.35$) & $0.00$ ($0.00$) & $-0.02$ ($0.03$) & $-0.15$ ($0.08$) & $0.05$ ($0.20$) & $0.01$ ($0.08$) & $-0.14$ ($0.10$) \\
\hline
\end{tabular}
\end{adjustbox}
\caption{Estimated ARX parameters and standard errors (between parenthesis) for scenario 1 (no influence of climate time series on simulated insect abundances), scenario 2 (influence of five climate time series on simulated insect abundances), and scenario 3 (influence of all climate time series on simulated insect abundances) with lags, $p = \{1, 3, 5\}$.}
\label{ARXParameters}
\end{table}

\begin{figure}[!ht]\centering
\includegraphics[width=1\textwidth]{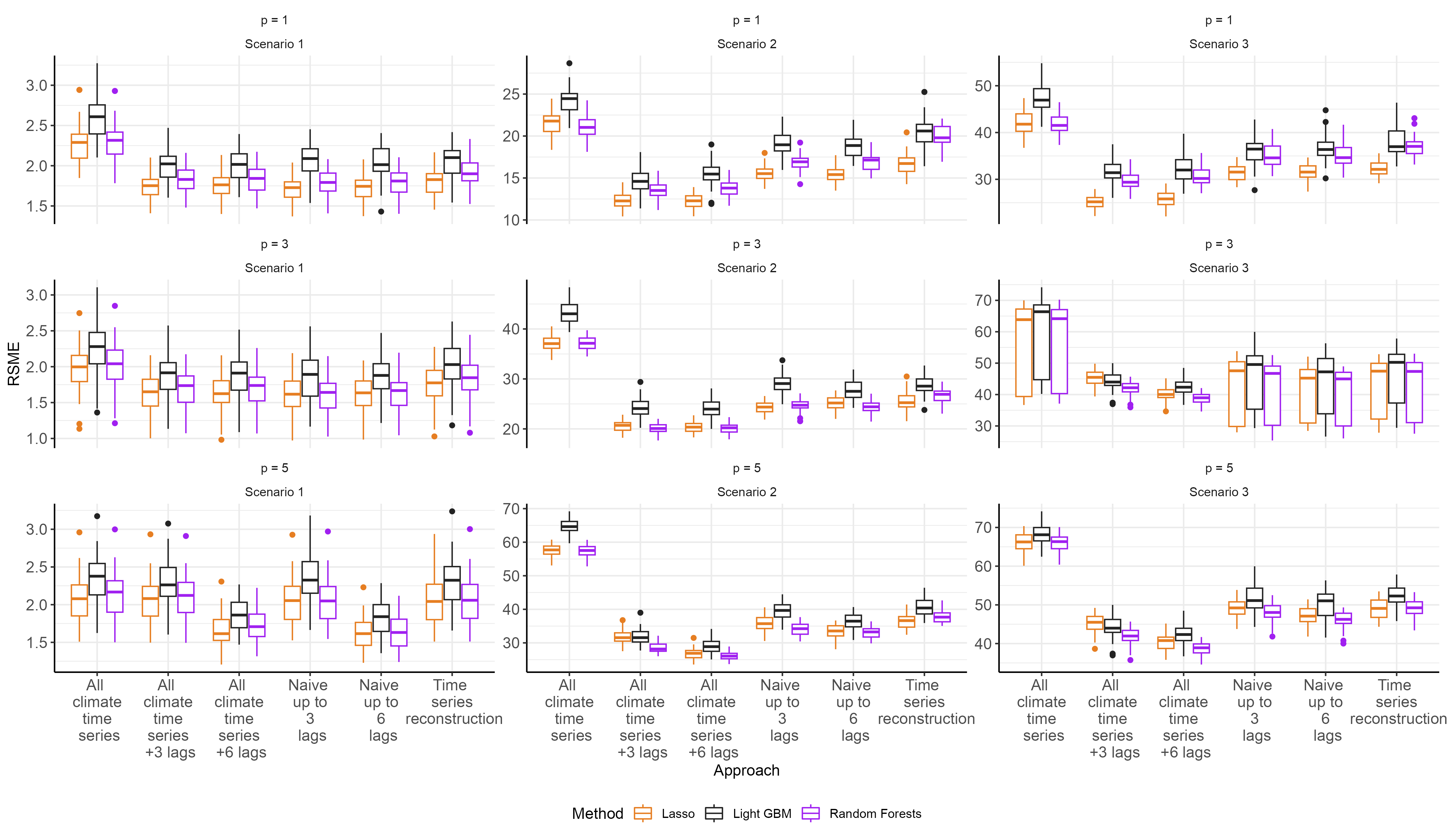}
\caption{Performance of the Random Forests, Lasso-regularised linear regression, and LightGBM algorithms for each approach (including all climate time series with no lags, time series reconstruction, taking the naive approach with up to $3$ or $6$-step lagged target series, and all climate times series with up to $3$ or $6$-step lagged target series) considering the simulation study where the initial number of training samples is $30$ and insect abundance is generated based on the \textbf{Poisson} ARX.}
\label{simulation_results_Poisson}
\end{figure}

\begin{figure}[!ht]\centering
\includegraphics[width=1\textwidth]{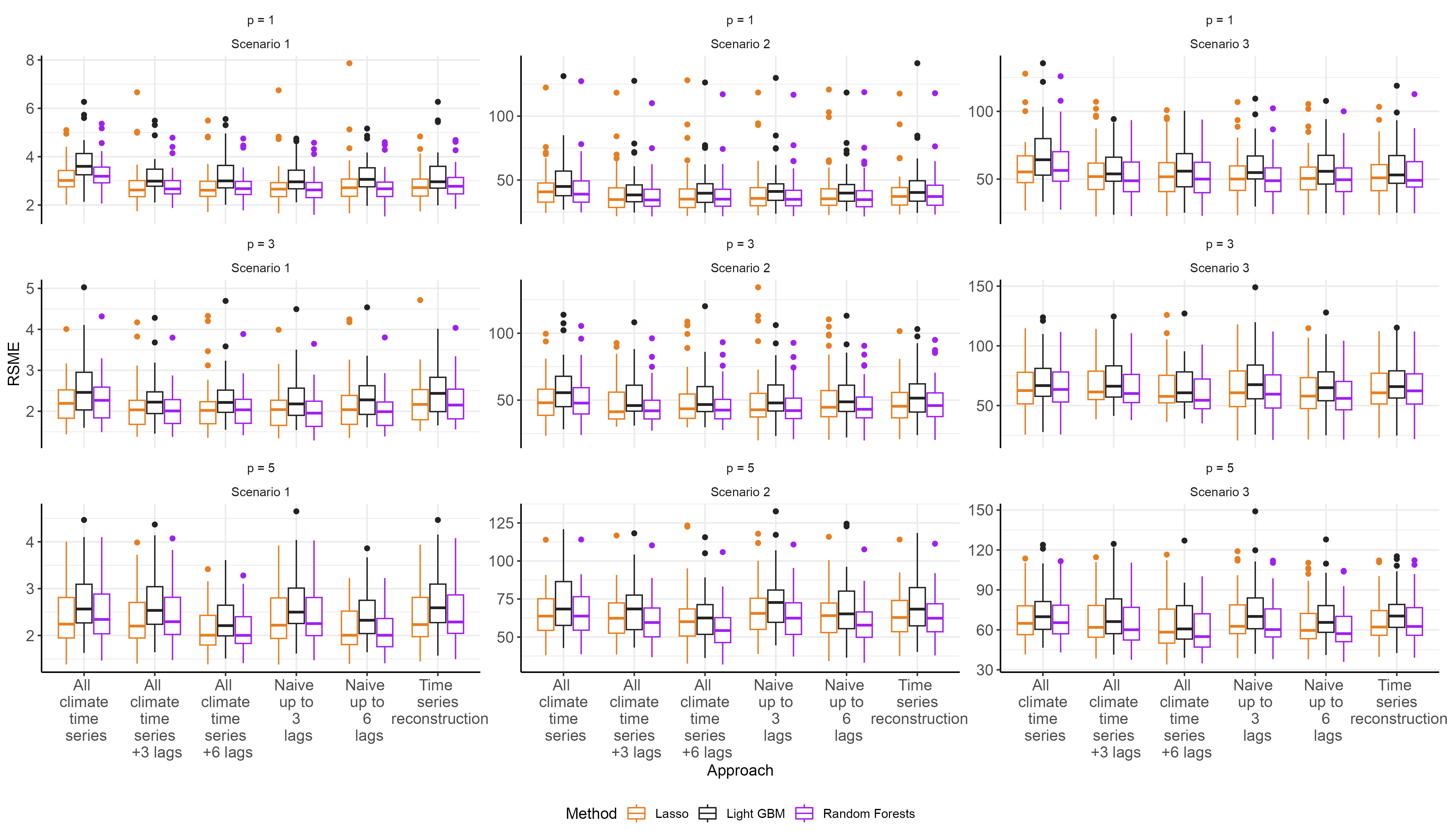}
\caption{Performance of the Random Forests, Lasso-regularised linear regression, and LightGBM regression algorithms for each approach (including all climate time series with no lags, time series reconstruction, taking the naive approach with up to $3$ or $6$-step lagged target series, and all climate times with up to $3$ or $6$-step lagged target series) considering the simulation study where the initial number of training samples is $30$ and insect abundance is generated based on the \textbf{negative binomial} ARX.}
\label{simulation_results_NBinomial}
\end{figure}

Considering all scenarios, approaches and learning algorithms, the obtained average RMSE for the Poisson ARX was $24.3$ with a standard deviation (sd) of $19.40$ for training the learning algorithms with $30$ initial samples and an average of $25.1$ ($\mbox{sd} = 20.70$) for $60$ initial samples. The obtained average RMSE for the negative binomial ARX was $39.1$ ($\mbox{sd} = 30.80$) for training the learning algorithms with $30$ initial samples and an average of $40.2$ ($\mbox{sd} = 32.40$) for $60$ initial samples. Given the slight difference between initial training samples on the forecasting performance, we solely present in Figure~\ref{simulation_results_Poisson} and Figure~\ref{simulation_results_NBinomial} the performance of the learning algorithms trained with $30$ initial training samples used to start the one-step ahead forecasting for Poisson and negative binomial ARX. 

In Figure~\ref{simulation_results_Poisson}, the performance of the learning algorithms for the simulation study is presented based on the Poisson ARX. Overall, the average RMSE for Random Forests, Lasso-regularised linear regression and LightGBM were, respectively, $23.7$ ($\mbox{sd} = 18.80$), $23.5$ ($\mbox{sd} = 19.00$) and $25.8$ ($\mbox{sd} = 20.20$), indicating that for the majority of the scenarios, values of $p$ and approaches the learning algorithms have similar performances. However, some scenarios, such as Scenario 2, Lasso-regularised linear regression and Random Forests, perform better than LightGBM. 

For scenario 1, where there is no influence of climate time series on the generation of insect abundances based on Poisson ARX, the average RMSE considering all values of $p$ for the proposed method, all climate time series with no lags, all climate time series with up to $3$ or $6$-lagged target series, and taking the naive approach with up to $3$ or $6$-lagged target series were, respectively, $1.97$ ($\mbox{sd} = 0.30$), $2.22$ ($\mbox{sd} = 0.34$), $1.91$ ($\mbox{sd} = 0.32$), $1.78$ ($\mbox{sd} = 0.25$), $1.89$ ($\mbox{sd} = 0.34$) and $1.75$ ($\mbox{sd} = 0.25$). For scenario 2, where there is an influence of five climate time series based on Poisson ARX, the average RMSE considering all values of $p$ for the proposed method, all climate time series with no lags, all climate time series with up to $3$ or $6$-lagged target series, and taking the naive approach with up to $3$ or $6$-lagged target series were, respectively, $28.1$ ($\mbox{sd} = 8.26$), $40.5$ ($\mbox{sd} = 15.70$), $21.9$ ($\mbox{sd} = 7.34$), $20.9$ ($\mbox{sd} = 5.88$), $26.5$ ($\mbox{sd} = 8.28$) and $25.7$ ($\mbox{sd} = 7.38$). For scenario 3, where there is an influence of all climate time series based on Poisson ARX, the average RMSE considering all values of $p$ for the proposed method, all climate time series with no lags, all climate time series with up to $3$ or $6$-lagged target series, and taking the naive approach with up to $3$ or $6$-lagged target series were, respectively, $43.3$ ($\mbox{sd} = 8.31$), $56.2$ ($\mbox{sd} = 12.2$), $38.7 $ ($\mbox{sd} = 7.67$), $36.8$ ($\mbox{sd} = 5.93$), $42.3$ ($\mbox{sd} = 8.81$) and $41.5$ ($\mbox{sd} = 7.96$). 

In Figure~\ref{simulation_results_NBinomial}, the performance of the learning algorithms for the simulation study is presented based on the negative binomial ARX. Overall, the average RMSE for Random Forests, Lasso-regularised linear regression and LightGBM were, respectively, $37.2$ ($\mbox{sd} = 28.90$), $38.5$ ($\mbox{sd} = 30.60$) and $41.7$ ($\mbox{sd} = 32.40$), indicating that for all scenarios, values of $p$ and approaches the learning algorithms have similar performances. For scenario 1, where there is no influence of climate time series on the generation of insect abundances based on negative binomial ARX, the average RMSE considering all lags for the proposed method, all climate time series with no lags, all climate time series with up to $3$ or $6$-lagged target series, and taking the naive approach with up to $3$ or $6$-lagged target series were, respectively, $2.60$ ($\mbox{sd} = 0.70$), $2.75$ ($\mbox{sd} = 0.80$), $2.51$ ($\mbox{sd} = 0.70$), $2.43$ ($\mbox{sd} = 0.69$), $2.49$ ($\mbox{sd} = 0.70$) and $2.44$ ($\mbox{sd} = 0.72$). 

For scenario 2, where there is an influence of five climate time series based on negative binomial ARX, the average RMSE considering all lags for the proposed method, all climate time series with no lags, all climate time series with up to $3$ or $6$-lagged target series, and taking the naive approach with up to $3$ or $6$-lagged target series were, respectively, $53.0$ ($\mbox{sd} = 19.60$), $56.0$ ($\mbox{sd} = 19.80$), $50.9$ ($\mbox{sd} = 19.00$), $50.2$ ($\mbox{sd} = 19.30$), $53.1$ ($\mbox{sd} = 21.20$) and $52.1$ ($\mbox{sd} = 20.20$). For scenario 3, where there is an influence of all climate time series based on negative binomial ARX, the average RMSE considering all lags for the proposed method, all climate time series with no lags, all climate time series with up to $3$ or $6$-lagged target series, and taking the naive approach with up to $3$ or $6$-lagged target series were, respectively, $62.4$ ($\mbox{sd} = 19.30$), $66.0$ ($\mbox{sd} = 19.70$), $62.6$ ($\mbox{sd} = 19.30$), $59.9$ ($\mbox{sd} = 18.50$), $62.9$ ($\mbox{sd} = 20.70$) and $60.4$ ($\mbox{sd} = 19.10$). 

\begin{figure}[!ht]\centering
\includegraphics[width=1\textwidth]{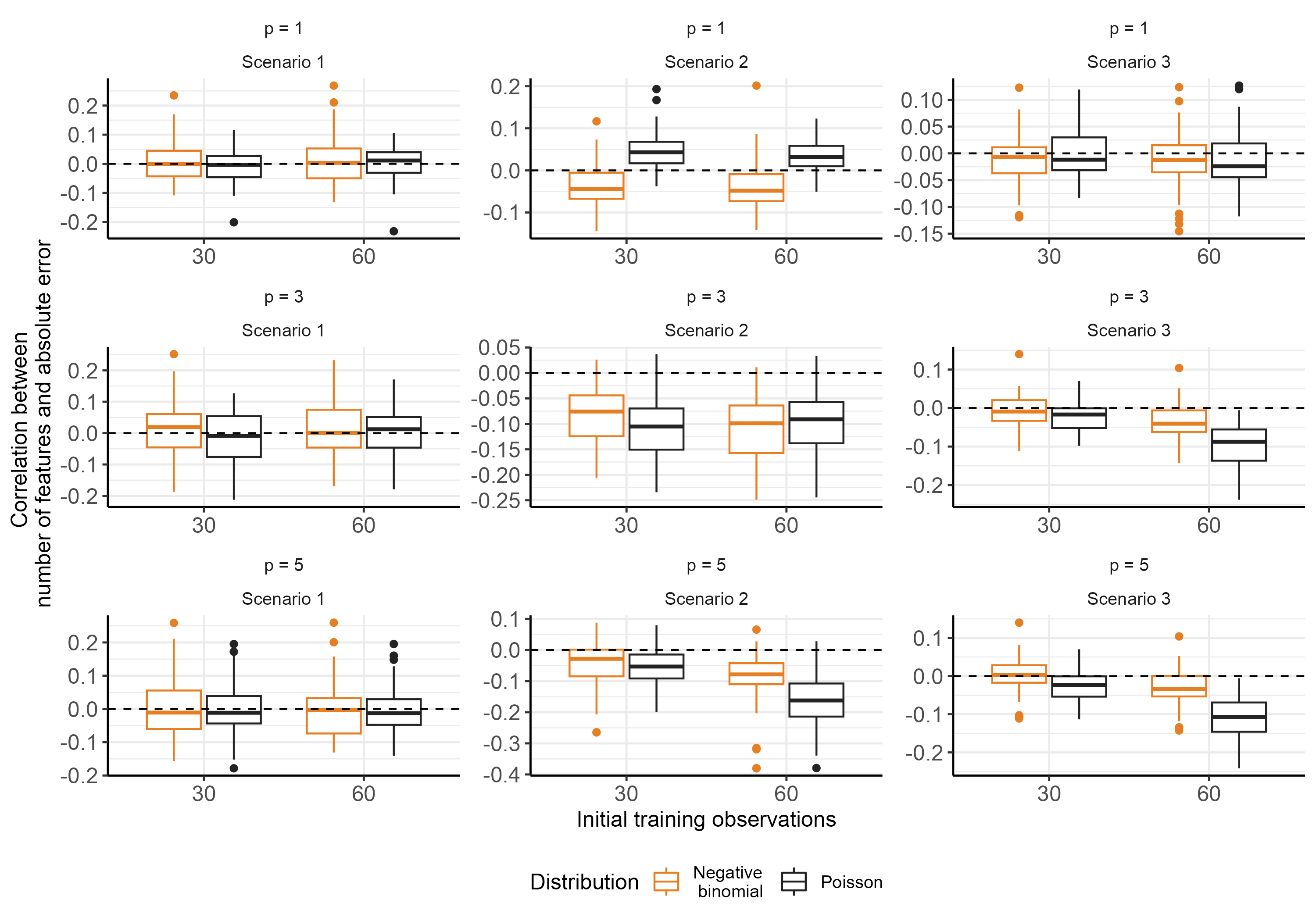}
\caption{Boxplots of the correlation between the Random Forests's forecasting absolute error and the number of selected features of $\mathcal{D}$ (the sum of the number of climate, target time series and their lags) per forecast by our approach for the simulated study. The correlations are presented for the case where the Random Forests algorithm was trained with $30$ and $60$ initial observation to start the one-step ahead forecasting. The dashed line indicates the correlation equal to zero.}
\label{simulation_study_selected_variables}
\end{figure}

Figure~\ref{simulation_study_selected_variables} shows the average correlation between the number of selected features of $\mathcal{D}$ (the sum of the number of climate, target time series and their lags) and the absolute prediction error using our approach with Random Forests to obtain forecasts. Overall, the percentage of negative correlations considering all scenarios and values of $p$ based on the Poisson ARX for $30$ and $60$ initial samples to start the one-step ahead forecasting were $63.9\%$ and $69.0\%$. Based on the negative binomial ARX for $30$ and $60$, initial samples to start the one-step ahead forecasting were $61.6\%$ and $71.3\%$. Also, the average correlations considering all scenarios and values of $p$ based on the Poisson ARX for $30$ and $60$ initial samples to start the one-step ahead forecasting were $-0.02$ ($\mbox{sd} = 0.07$) and $-0.05$ ($\mbox{sd} = 0.09$). Based on the negative binomial ARX for $30$ and $60$ initial samples to start the one-step ahead forecasting were $-0.02$ ($\mbox{sd} = 0.07$) and $-0.03$ ($\mbox{sd} = 0.08$). It indicates that the correlation is close to zero, and most correlations are negative. 

For scenario 1, the percentage of negative correlations considering all values of $p$ based on the Poisson ARX for $30$ and $60$ initial samples to start the one-step ahead forecasting were $55.5\%$ and $48.6\%$. Based on the negative binomial ARX for $30$ and $60$ initial samples to start the one-step ahead forecasting were $49.3\%$ and $50.7\%$. Also, the average correlations considering all values of $p$ based on the Poisson ARX for $30$ and $60$ initial samples to start the one-step ahead forecasting were $-0.01$ ($\mbox{sd} = 0.07$) and $0.00$ ($\mbox{sd} = 0.07$). Based on the negative binomial ARX for $30$ and $60$ initial samples to start the one-step ahead forecasting were $0.01$ ($\mbox{sd} = 0.08$) and $0.01$ ($\mbox{sd} = 0.08$). 

For scenario 2, the percentage of negative correlations considering all values of $p$ based on the Poisson ARX for $30$ and $60$ initial samples to start the one-step ahead forecasting were $66.0\%$ and $70.8\%$. Based on the negative binomial ARX for $30$ and $60$ initial samples to start the one-step ahead forecasting were $83.3\%$ and $91.3\%$. In addition, the average correlations considering all values of $p$ based on the Poisson ARX for $30$ and $60$ initial samples to start the one-step ahead forecasting were $-0.04$ ($\mbox{sd} = 0.08$) and $-0.08$ ($\mbox{sd} = 0.10$). Based on the negative binomial ARX for $30$ and $60$ initial samples to start the one-step ahead forecasting were $-0.05$ ($\mbox{sd} = 0.06$) and $-0.08$ ($\mbox{sd} = 0.08$). 

For scenario 3, the percentage of negative correlations considering all values of $p$ based on the Poisson ARX for $30$ and $60$ initial samples to start the one-step ahead forecasting were $70.0\%$ and $87.5\%$. Based on the negative binomial ARX for $30$ and $60$ initial samples to start the one-step ahead forecasting were $52.0\%$ and $72.0\%$. The average correlations considering all values of $p$ based on the Poisson ARX for $30$ and $60$ initial samples to start the one-step ahead forecasting were $-0.01$ ($\mbox{sd} = 0.04$) and $-0.07$ ($\mbox{sd} = 0.07$). Based on the negative binomial ARX for $30$ and $60$ initial samples to start the one-step ahead forecasting were $0.00$ ($\mbox{sd} = 0.05$) and $-0.03$ ($\mbox{sd} = 0.05$).

\section{Discussion}

We presented a new approach for reconstructing time series dependencies using Takens' embedding theorem and Granger's causality. The approach can automatically select target and climate time series, including their lags, and we propose using machine learning algorithms that "learn" from the reconstructed time series to forecast insect abundance. We applied our proposed methods to two different datasets of insect time series and climate covariates associated with every observation of insect abundance. Also, a simulation study was introduced to explore the novel approach. The case study illustrates that the proposed approach is competitive compared with the other approaches presented in this paper. Also, the correlations presented by Figure~\ref{case_study_features} bring insights concerning the effect of the number of selected features on the forecasting performance, indicating that more features do not negatively impact the overall performance. The approaches using a combination of climate and the target time series performed better than those using solely the climate time series. It indicates the importance of using the target time series as features for the machine learning algorithms for forecasting insect abundance. 

Furthermore, the simulation study based on Poisson ARX emphasised the importance of adding the target time series as features for the machine learning methods. When the importance of the climate time series was highlighted in scenarios 2 and 3, the poor performance of the approach that uses all climate time series solely with no lags to predict insect abundance is more evident compared to the others. For the simulation study based on negative binomial ARX, most approaches obtained similar performances due to the super dispersion provided by the negative binomial distribution. It indicates that when the association of climate and the target time series are blurred by super dispersion, the implemented approaches perform similarly. 

Figure~\ref{simulation_study_selected_variables} highlights the effect of the number of selected features of $\mathcal{D}$ (the sum of the number of climate, target time series and their lags) on the forecasting performance of the proposed approach based on the Random Forests' forecasting absolute error. Our results showed that most simulation scenarios' correlations are closer to zero or negative. We only observed negative and positive correlations with relatively higher intensity in scenario 2. It indicates the importance of selecting the lags and climate time series for forecasting insect abundances. Therefore, our approach provides selection criteria for lags and climate time series that impact the forecast performance of the Random Forests algorithm in the minority of the simulated scenarios. The impact is translated into selecting fewer climate time series in scenarios where they present an influence on the simulated insect abundance.

Our results indicate that the proposed method is competitive with the other approaches to applying machine learning to forecast insect abundances. Other researchers have proposed causal discovery methods for time series analysis based on different methodologies~\citep{eichler2010granger, eichler2013, runge2019detecting, glymour2019review, assaad2022entropy, yuan2022data, runge2023}. However, combining machine learning algorithms and causal discovery based on Takens' embedding theory targeting forecasting presents a novel contribution to entomologists applying a learning-based algorithm to forecasting insect abundance. Thus, our work will be a basis for developing new techniques in this domain.

\section{Conclusion}

The proposed approach presented a competitive performance with commonly used approaches to forecast insect abundance in terms of predictive power. It shows the feasibility of applying these techniques to forecasting insect pest abundance, and therefore, this study constitutes a basis for developing new techniques to predict insect outbreaks. Also, the feature selection of our method does not negatively influence the forecasting capability of the Random Forests algorithm, indicating that even if a climate time series does not influence the target time series, our methodology can select features that maintain a competitive forecasting performance compared to approaches presented in our study.

\section{Acknowledgments}

This publication has emanated from research conducted with the financial support of Science Foundation Ireland under Grant 18/CRT/6049. The opinions, findings and conclusions or recommendations expressed in this material are those of the authors and do not necessarily reflect the views of the funding agencies.

\section{Declarations}
~~~~
\textbf{Ethical Approval} Not applicable.

\textbf{Competing interests} Not applicable.

\textbf{Authors’ contributions} G.R.P., W.A.C.G. and R.F.M. conceived and designed the research. E.E. and D.L. collected the data. C.M., E.E. and D.L. provided insights into the discussion of results. G.R.P. and R.F.M. created the methodology and analysed the data. G.R.P., R.F.M., C.M. and R.A.M. led the writing of the manuscript. All authors contributed to the overall writing. G.R.P. and R.A.M. created the R implementation of the methodology.

\textbf{Funding} Science Foundation Ireland under Grant 18/CRT/6049


\bibliographystyle{apalike}
\bibliography{ref}

\end{document}